\documentclass[fleqn,usenatbib]{mnras}

\usepackage{newtxtext,newtxmath}

\usepackage[T1]{fontenc}

\DeclareRobustCommand{\VAN}[3]{#2}
\let\VANthebibliography\thebibliography
\def\thebibliography{\DeclareRobustCommand{\VAN}[3]{##3}\VANthebibliography}

\usepackage{graphicx}	
\usepackage{amsmath}	
\usepackage{tabu,booktabs}
\usepackage{xcolor}
\usepackage[english]{babel}
\usepackage{graphicx}
\usepackage{xspace}

\usepackage{longtable}
\usepackage{threeparttablex}
\usepackage{array}

\title[NIHAO XXVII: Crossing the green valley]{NIHAO XXVII: Crossing the green valley}

\author[M. Blank, A. V. Macci\`o, X. Kang, K. L. Dixon, N. H. Soliman]{
Marvin Blank,$^{1,2,3}$\thanks{E-mail: marvin.blank@nyu.edu (MB)}, Andrea V. Macci\`o$^{1,2,4}$, Xi Kang$^{5,6}$, Keri L. Dixon,$^{1,2}$ and Nadine H. Soliman$^{7,1}$\\
$^{1}$New York University Abu Dhabi, PO Box 129188, Saadiyat Island, Abu Dhabi, United Arab Emirates\\
$^{2}$Center for Astro, Particle and Planetary Physics (CAP$^3$), New York University Abu Dhabi \\
$^{3}$Institut f\"{u}r Theoretische Physik und Astrophysik, Christian-Albrechts-Universit\"{a}t zu Kiel, Leibnizstr. 15, D-24118 Kiel, Germany\\
$^{4}$Max Planck Institut f\"{u}r Astronomie, K\"{o}nigstuhl 17, D-69117 Heidelberg, Germany\\
$^{5}$Zhejiang University-Purple Mountain Observatory Joint Research Center for Astronomy, Zhejiang University, Hangzhou 310027, China\\
$^{6}$Purple Mountain Observatory, No 8 Yuanhua Road, Nanjing 210034, China\\
$^{7}$California Institute of Technology, Pasadena, CA 91125, USA
}

\date{Accepted XXX. Received YYY; in original form ZZZ}

\pubyear{2021}

\begin{document}
\label{firstpage}
\pagerange{\pageref{firstpage}--\pageref{lastpage}}
\maketitle

\begin{abstract}

The transition of high-mass galaxies from being blue and star forming to being red and dead is a crucial step in galaxy evolution, yet not fully understood.
In this work, we use the NIHAO suite of galaxy simulations to investigate the relation between the transition time through the green valley and other galaxy properties.
The typical green valley crossing time of our galaxies is approximately 400 Myr, somewhat shorter than observational estimates.
The crossing of the green valley is triggered by the onset of AGN feedback and the subsequent shut down of star formation. 
Interestingly the time spent in the green valley is not related to any other galaxy properties, such as stellar age or metallicity, or the time at which the star formation quenching takes place.
The crossing time is set by two main contributions: the ageing of the current stellar population and the residual star formation in the green valley. These effects are of comparable magnitude, while major and minor mergers have a negligible contribution.
Most interestingly, we find the time that a galaxy spends to travel through the green valley is twice the $e$-folding time of the star formation quenching.
This result is stable against galaxy properties and the exact numerical implementation of AGN feedback in the simulation.
Assuming a typical crossing time of about one Gyr inferred from observations, our results imply that any mechanism or process aiming to quench star formation, must do it on a typical timescale of 500 Myr. 

\end{abstract}




\section{Introduction}\label{sec:intro}

Several observations indicate a bimodality in the colour-magnitude diagram of galaxies,
separating galaxies into the {\it red sequence} and the {\it blue cloud} \citep{2001_Strateva_Ivezic_Knapp,2005_Kang_Jing_Mo}.
The former consists of early-type galaxies with little or no star formation, the latter of late-type galaxies with higher star formation rates.
These two populations are separated by the {\it green valley} (GV), which is very sparsely populated \citep{2006_Baldry_Balogh_Bower,2006_Driver_Allen_Graham,2007_Wyder_Martin_Schiminovich}.
Galaxies evolve from the blue cloud to the red sequence due to star formation quenching
\citep{2021_Blank_Meier_Maccio,2013_Fang_Faber_Koo,2007_Schawinski_Thomas_Sarzi}, most likely due to AGN (active galactic nucleus) feedback \citep{2019_Blank_Maccio_Dutton,2006_Croton_Springel_White,2005_DiMatteo_Springel_Hernquist,2005_Murray_Quataert_Thompson}.
Much of the literature concludes that few galaxies lie in the GV, because they cross it relatively quickly compared to cosmic timescales \citep{2019_Wright_Lagos_Davies, 2014_Schawinski_Urry_Simmons}.
There are different methods to define the GV, e.g. by specifying a color cut as a function of stellar mass and redshift \citep{2014_Schawinski_Urry_Simmons,2018_Coenda_Martinez_Muriel}, or by using the 4000 \AA \, break strength \citet{2020_Angthopo_Ferreras_Silk}.

Actually, most observations do not measure the GV crossing time, but the $e$-folding timescale of an (assumed) exponentially declining star formation rate.
\citet{2013_Wetzel_Tinker_Conroy} observe that for satellites falling into a cluster, star formation quenches after an $e$-folding timescale of <~0.8~Gyr.
\citet{2014_Schawinski_Urry_Simmons} find that spheroids quench faster than discs, and \citet{2018_Smethurst_Masters_Lintott} conclude that slow rotators quench more rapidly (<~1~Gyr) than fast rotators.
\citet{2018_Bremer_Phillipps_Kelvin} find quenching times of $\sim$1-2~Gyr, which do not depend on environment; however, the onset of quenching is dependent on environment.
\citet{2016_Carollo_Cibinel_Lilly} investigate how the quenched galaxy fraction depends on galaxy and environment properties.

Several theoretical studies have investigated either quenching times or GV crossing times by means of numerical simulations:
\citet{2019_Wright_Lagos_Davies} investigate quenching timescales in the EAGLE simulations \citep{2015_Schaye_Crain_Bower}. The authors find that high-mass centrals have a quenching time of <~2~Gyr, low-mass centrals quench due to stellar feedback in about 4~Gyr, and satellites rapidly quench in 2~Gyr due to ram pressure stripping.
These quenching times show a strong correlation with gas (<~30~kpc) fraction (though not total gas fraction) and with the ratio between stellar mass and halo mass, while there is no clear correlation with stellar mass.
\citet{2019_Correa_Schaye_Trayford} examine the origin of the red sequence in the EAGLE simulations and find that the GV crossing time depends weakly on morphology: elliptical galaxies cross in about 1~Gyr and disc galaxies in about 1.5~Gyr.
For satellites, the GV crossing time is positively correlated with the ratio between stellar mass and halo mass, whereas for centrals, there is no such correlation.
However, the time galaxies become green is correlated with the time at which the black hole (BH) accretion rate peaks.
\citet{2018_Nelson_Pillepich_Springel} use the Illustris TNG simulations \citep{2018_Pillepich_Springel_Nelson} to examine the galaxy colour bimodality and find GV crossing times of $\sim$1.6~Gyr, which drop for increasingly more massive galaxies.
The authors also conclude that quenching is mostly caused by BHs. This finding is supported by \citet{2017_Sparre_Springel}, where quenching is shown only to occur if AGN feedback is sufficiently strong.
\citet{2017_Hahn_Tinker_Wetzel} investigate star formation quenching timescales in centrals, finding $e$-folding times of 0.5-1.5~Gyr and more massive galaxies quenching faster.
Satellite galaxies experience external quenching, whereas central galaxies are quenched
internally, possibly by strangulation, i.e. gas depletion.
\citet{2016_Trayford_Theuns_Bower} examine the evolution of galaxy colour in the EAGLE simulations. These simulated galaxies turn red, because either they become satellites or experience AGN feedback, with GV crossing times of $\sim$2~Gyr, independent of galaxy mass and cause of quenching.

In this paper, we calculate the GV crossing times of galaxies from the NIHAO project \citep[Numerical Investigation of a Hundred Astrophysical Objects][]{2015_Wang_Dutton_Stinson,2019_Blank_Maccio_Dutton} and investigate whether these timescales correlate with other galaxy quantities.
In section~\ref{sec:nihao}, we introduce the NIHAO simulations, and in section~\ref{sec:measure}, we explain how we measure the relevant quantities, like the GV boundaries.
In section~\ref{sec:results}, we present our results, in section~\ref{sec:outlook} we elaborate on possible future research, and we summarize our findings in section~\ref{sec:summary}.

\section{The NIHAO simulations}\label{sec:nihao}

The NIHAO suite of galaxy simulations consists of more than 150 zoom-in simulations of central galaxies, introduced by \citet{2015_Wang_Dutton_Stinson} and extended by \citet{2019_Blank_Maccio_Dutton}.
Their initial conditions are derived from cosmological dark matter simulations with $400^3$ particles and with box sizes of 60, 20, and 15 $\rmn{Mpc} \, \rmn{h}^{-1}$.
These boxes are simulated until redshift zero, then haloes are extracted and resimulated individually with gas particles and a higher resolution.
The zoom-in simulations have resolutions that resolve the mass profile at $\leq$ 1~per~cent of the virial radius and that each galaxy contains about $10^6$ particles.

This choice gives dark matter particles with softening lengths of 116 to 931~pc and masses of $3.4 \times 10^3$ to $1.7 \times 10^6 \, \rmn{M}_{\sun}$.
The ratio of dark and gas particle mass is set to the cosmological mass ratio of dark matter and baryons, $\Omega_{\rmn{DM}}/\Omega_{\rmn{b}} = 5.48$;
the ratio of dark and gas particle softening length is  $(\Omega_{\rmn{DM}}/\Omega_{\rmn{b}})^{1/2} = 2.34$.
We use cosmological parameters from the \citet{2014_Planck_Collaboration}, i.e.a flat LCDM universe.
NIHAO is simulated with the TreeSPH code {\sc Gasoline2} \citep{2017_Wadsley_Keller_Quinn}.
The gas cools via hydrogen, helium, and various metal-lines in a uniform UV ionizing background \citep{2010_Shen_Wadsley_Stinson}, including UV background heating \citep{2012_Haardt_Madau}, photoionization, and Compton cooling.
For star formation, gas particles that exceed a temperature and density threshold ($n > 10.3 \, \rmn{cm}^{-3}$, $T < 15000 \, \rmn{K}$) form stars with a rate of $\dot{M_{\star}} = c_{\star} M_{\rmn{gas}} t_{\rmn{dyn}}^{-1}$, where $c_{\star}=0.1$ is the star formation efficiency, $t_{\rmn{dyn}} = (4 \uppi G \rho)^{-1/2}$ is the gas particle's dynamical time and $\rho$ its density, and $M_{\rmn{gas}}$ its mass.
We use the blastwave formalism of \citet{2006_Stinson_Seth_Katz} to model supernova feedback; here, star particles with $8 < M_{\star}/M_{\sun} < 40$ inject thermal energy and metals into surrounding gas particles 4~Myr after their formation\footnote{in star forming regions the time step is around 50,000 yr}, and after that time, the cooling of these gas particles is switched off for $\sim$~30~Myr.
Before going supernova, star particles provide \lq early stellar feedback\rq \,\citep{2013_Stinson_Brook_Maccio}; here, 13~per~cent of the total stellar flux of $2 \times 10^{50} \, \rmn{erg} \, \rmn{M}_{\sun}^{-1}$ is transferred to the surrounding gas as thermal energy.
The free parameters of the stellar and supernova feedback model have been chosen to reproduce the $M_{\star}$-$M_{{200}}$ relation for one Milky Way-like galaxy at $z=0$.
For further details on the NIHAO project, see \citet{2015_Wang_Dutton_Stinson}.

BHs with a seed mass of $10^{5}\,\rmn{M}_{\sun}$ form in the center of haloes that exceed a threshold mass of $5 \times 10^{10}\,\rmn{M}_{\sun}$. At every major time step, the BH is relocated to the dark matter particle within ten softening lengths that has the lowest gravitational potential.
BHs accrete gas from neighbouring gas particles with the Bondi-Hoyle-Lyttleton accretion rate \citep{1939_Hoyle_Lyttleton, 1944_Bondi_Hoyle, 1952_Bondi}, which is limited by the Eddington rate \citep{1921_Eddington}.
For BH feedback, the gas receives an energy per time of $\dot{E} = \epsilon \dot{M}_{\rmn{BH}} c^2$, with $\epsilon = 0.005$, BH accretion rate $\dot{M}_{\rmn{BH}}$, and speed of light $c$. This energy is distributed kernel weighted among the 50 nearest gas particles. See \citet{2019_Blank_Maccio_Dutton} for more details regarding the implementation of BH physics.

NIHAO reproduces several galaxy properties for halo masses of $M_{200} \leq 2 \times 10^{12}\,\rmn{M}_{\sun}$, e.g., the stellar mass versus halo mass relation \citep{2015_Wang_Dutton_Stinson}, the galaxy velocity function \citep{2016_Maccio_Udresco_Dutton}, the Tully-Fisher relation \citep{2017_Dutton_Obreja_Wang}, the rotation curves of dwarf galaxies \citep{2018_SantosSantos_DiCintio_Brook}, the stellar mass versus BH mass relation \citep{2019_Blank_Maccio_Dutton}, and the star formation rate versus stellar mass relation \citep{2021_Blank_Meier_Maccio}.

In this paper, we mainly use the 52 NIHAO galaxies from \citet{2019_Blank_Maccio_Dutton} with BH physics, which have $z=0$ halo masses ranging from $10^{12}\,\rmn{M}_{\sun}$ to $4 \times 10^{13}\,\rmn{M}_{\sun}$.
In section \ref{sec:measure}, for the definition of the GV, we additionally use 51 of the classical NIHAO galaxies without BHs from \citet{2015_Wang_Dutton_Stinson}.

\section{Definition of blue cloud, green valley, red sequence, and other quantities}\label{sec:measure}

To calculate the galaxy luminosities in the U and R band, we use the Binary Population and Spectral Synthesis (BPASS) models (v2.2.1) as described in \citet{2017_Eldridge_Stanway_Xiao} and \citet{2018_Stanway_Eldridge}.
As for dust contribution, \citet{2015_Trayford_Theuns_Bower} show that intrinsic, dust-free U-R colours are in agreement with observations. 
Furthermore, \citet{2018_Nelson_Pillepich_Springel} compare dust and no-dust models; significant changes occur only for very blue galaxies and thus are not relevant for investigating the GV.
Following \citet{2016_Dubois_Peirani_Pichon}, \citet{2019_Correa_Schaye_Trayford}, and \citet{2019_Wright_Lagos_Davies}, we do not account for dust in this work.
The exact details of the luminosity and colour calculation do not significantly affect the crossing times. Using e.g. the Padova simple stellar populations from \citet{2008_Marigo_Girardi_Bressan} and \citet{2010_Girardi_Williams_Gilbert} gives different absolute values of the colour boundaries, but about the same GV witdh, and thus does not change the GV crossing times significantly.

\begin{figure}
	\includegraphics[width=\columnwidth]{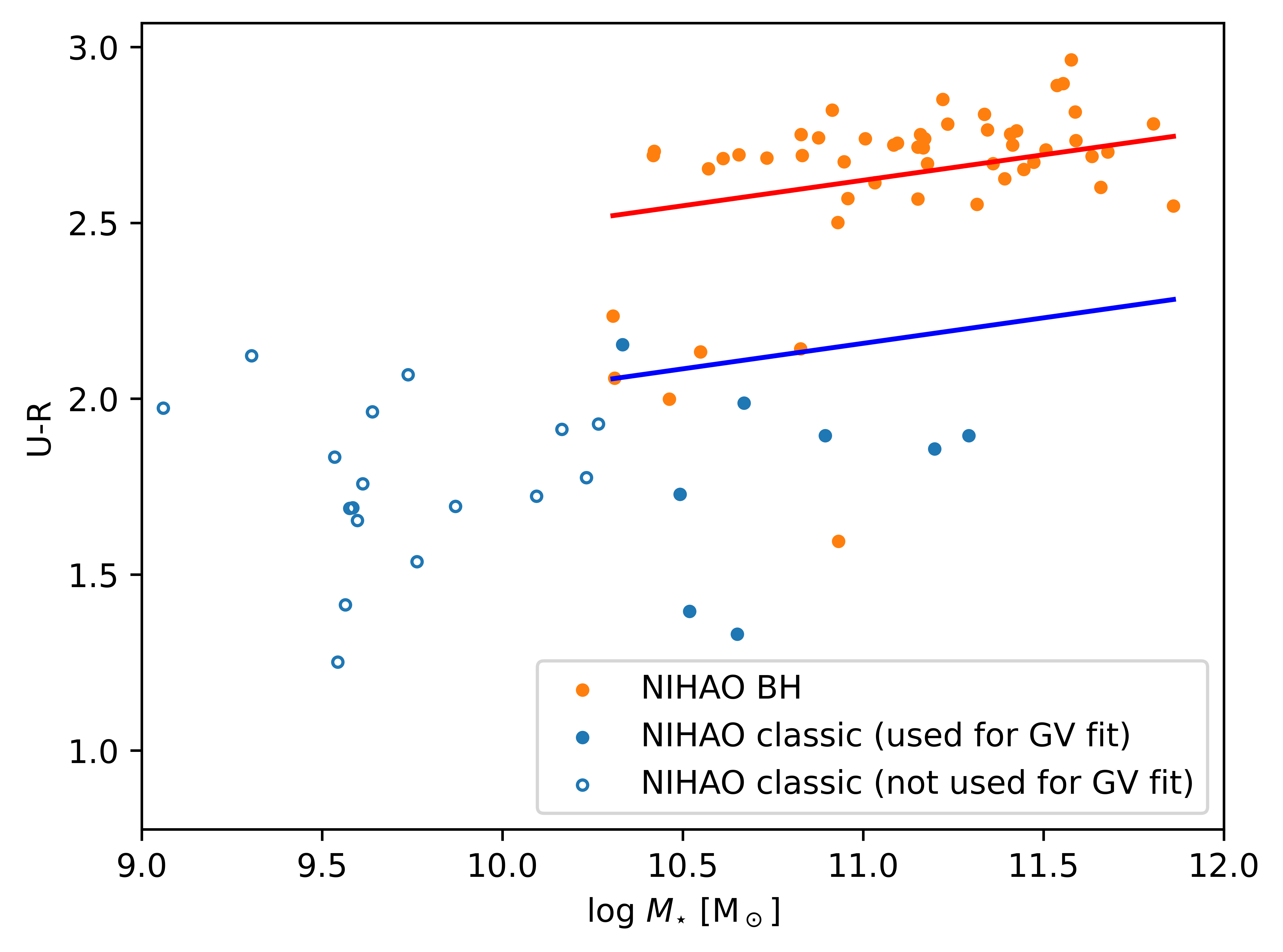}
    \caption{U-R colour versus stellar mass at redshift zero for the classical NIHAO galaxies (blue circles) and NIHAO galaxies with BHs (orange circles). The filled circles are galaxies used for the GV calculation, the empty circles are galaxies not used for the GV calculation.
    The red and blue line are the upper and lower boundary of the GV.}
    \label{fig:mag_mstar}
\end{figure}

Fig.~\ref{fig:mag_mstar} shows the galaxy colour versus stellar mass for the classical NIHAO galaxies \citep{2015_Wang_Dutton_Stinson} and the NIHAO galaxies with BHs \citep{2019_Blank_Maccio_Dutton}.
The combination of both samples clearly shows a bimodality in the galaxy distribution.
The red and blue lines denote the upper and lower boundary of the GV, respectively, which are a function of time and stellar mass.

To calculate the GV boundaries we use an adapted version of the machine learning {\it k-clustering} algorithm:
We assume that the colours $C_{\rmn{b,r}}$ of the blue (index `b') and red (index `r') group are functions of time $t$ and stellar mass $M_{\rmn{\star}}$ and are represented by a plane
\begin{equation}
    C_{\rmn{b,r}}(t,M_{\rmn{\star}}) = a \left( t / \rmn{Gyr} \right) + b \log \left( M_{\rmn{\star}} / \rmn{M}_{\sun} \right) + c_{\rmn{b,r}} \,,
    \label{eq:plane}
\end{equation}
where $a$, $b$, $c_{\rmn{b}}$ and $c_{\rmn{r}}$ are the fitting parameters.
Note that both, the red and the blue plane, have the same parameters $a$ and $b$. Furthermore we do not include a mixed term $\sim t \, \log M_{\rmn{\star}}$. Doing so would change the GV crossing times by only a few per cent, thus we omit these terms to reduce the number of free parameters of our model.
We initially choose a random representation of each plane.
We then
(i) assign each data point (which is a galaxy U-R colour at a specific time step with a specific stellar mass) to the blue group if it is closer to the blue plane, and to the red group if the data point is closer to the red plane,
(ii) attain a new representation of the planes $C_{\rmn{b,r}}$ by fitting eq.~\ref{eq:plane} to the blue and red group, respectively.
Repeating steps (i) and (ii) will obtain an ever better division of the data points into a blue and red group, we stop this process when no data point changes its group in step (i).
We only include data points with $t > 2~\rmn{Gyr}$, as no galaxy is red before that time.  
We furthermore only use classical NIHAO galaxies with a stellar mass of at least the smallest stellar mass of the NIHAO BH galaxies in each time step, because galaxies with a lower stellar mass do not have a counterpart with BH in the red sequence as seen in Fig.\,\ref{fig:mag_mstar}.
We then calculate the scatter $\epsilon_{\rmn{b,r}}$ of each plane such that 68 per cent of all data points of the respective group are included within the range $C_{\rmn{b,r}} \pm \epsilon_{\rmn{b,r}}$.
The lower (blue) boundary of the GV is then defined as $C_{\rmn{b}}(t,M_{\rmn{\star}}) + \epsilon_{\rmn{b}}$ and the upper (red) boundary as $C_{\rmn{r}}(t,M_{\rmn{\star}}) - \epsilon_{\rmn{r}}$. 
We obtain values of $a=0.092$, $b=0.055$, $c_{\rmn{b}}=0.144$,  $c_{\rmn{r}}=1.015$, $\epsilon_{\rmn{b}}=0.222$ and $\epsilon_{\rmn{r}}=0.158$.

These planes generally provide a good fit to both, the red and blue populations. Only at a few redshifts the fit might not match the galaxy colors too well, as seen e.g. in the blue boundary for $z=0$ in Fig.\,\ref{fig:mag_mstar}.
Our definition of the GV gives a GV width of about 0.5, and is thus in agreement with other works, e.g. \citet{2014_Schawinski_Urry_Simmons, 2016_Trayford_Theuns_Bower}. The absolute values of the red and blue boundaries is different though, but this does not significantly affect the GV crossing times as mentioned above.

The star formation rate (SFR) at a specific time is calculated as the mass of all star particles within 20~per~cent of the galaxy's virial radius that have formed in the last 200\,Myr, divided by 200\,Myr.
If no star particle forms within the last 200\,Myr, we estimate a lower limit of the SFR by dividing the gas particle mass by 200\,Myr.

\section{Results}\label{sec:results}

\subsection{Overview and individual examples}

From \citet{2019_Blank_Maccio_Dutton}, we have 52 galaxies ($M_{\mathrm{halo}} > 7 \times 10^{11}\,\mathrm{M}_{\sun}$) with AGN feedback, shown as red symbols in Fig.~\ref{fig:mag_mstar}.
Of these galaxies, 50 are initially blue, and two are initially green. At redshift zero, 37 galaxies are red, twelve are green and three are blue.
The vast majority of the galaxies cross the GV at least once: 39 galaxies cross once, seven galaxies cross twice, and two cross three times, and four do not cross the GV within a cosmological time.

Our results show that SF quenching is needed in order to cross the GV, which is provided by AGN feedback. The classical NIHAO galaxies do not include black holes, and thus do not have AGN feedback. However, our previous works \citep{2015_Wang_Dutton_Stinson} show that low mass galaxies, like NIHAO classic galaxies, are not affected by AGN anyway, thus even with black holes they would probably not cross the GV.
This is evident in Fig.\,\ref{fig:mag_mstar} by some of the NIHAO BH galaxies, which have a low mass and do not cross the GV.
Thus the division of galaxies into a blue cloud and a red sequence is mostly between galaxies of high mass that experience AGN feedback, and galaxies of low mass where AGN do not show strong feedback. Thus in our case the division is mostly between NIHAO classic and NIHAO BH galaxies.

\begin{figure*}
	\includegraphics[width=\textwidth]{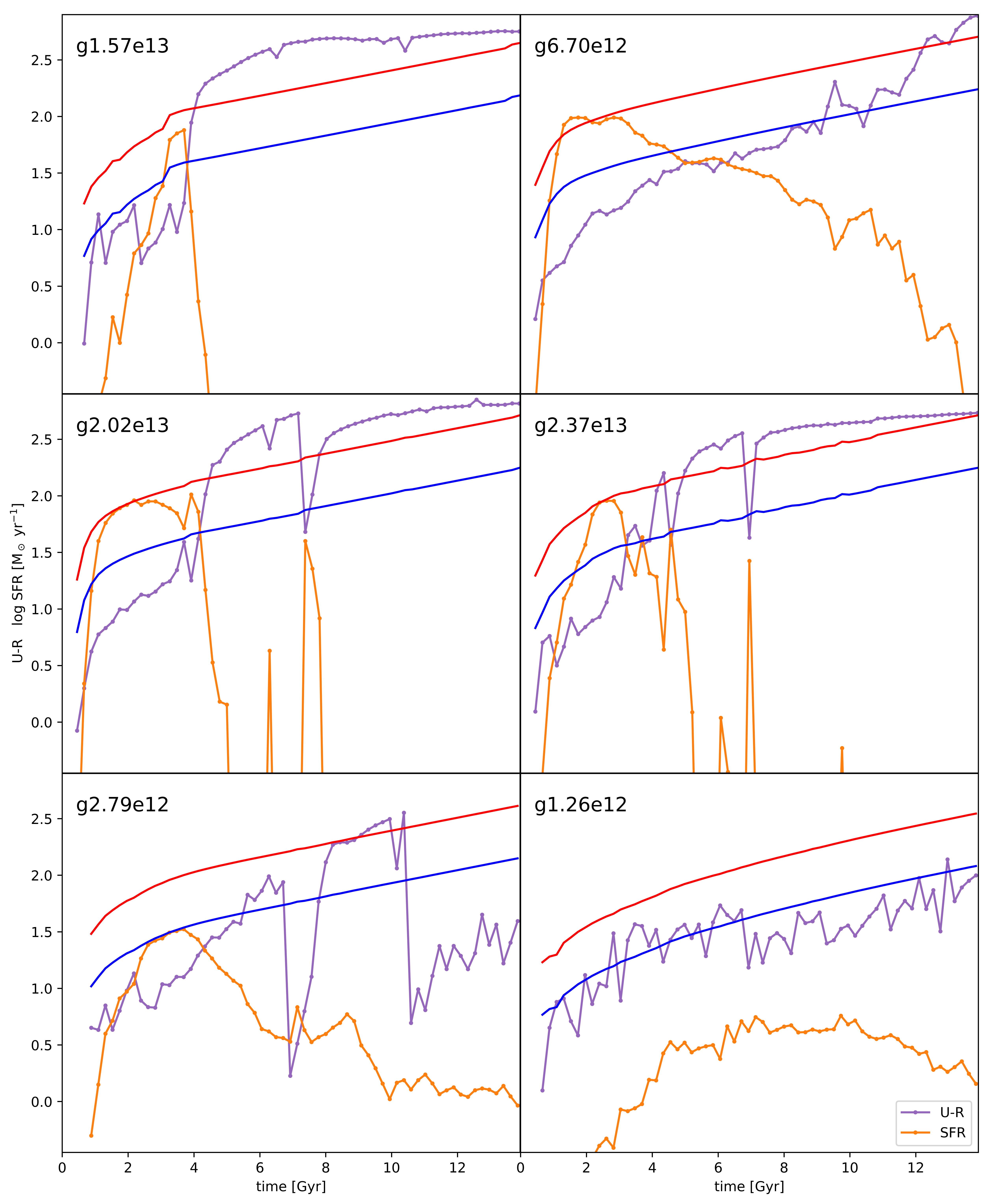}
    \caption{U-R colour (purple) and SFR (orange) versus time for six galaxies. The red and blue lines are the upper and lower boundary of the GV, respectively.
    The upper panel shows two `prototype' galaxies that cross the GV once in an orderly fashion. The middle and lower panels show galaxies with a more convoluted evolution.}
    \label{fig:mag_time_tracks_comp}
\end{figure*}

Fig.~\ref{fig:mag_time_tracks_comp} shows the colour and SFR of six galaxies versus time to demonstrate the variety of GV crossing times and the physical processes that can affect the crossing of the GV.
Galaxy g1.57e13 experiences a starburst starting at 3\,Gyr, which is abruptly quenched at about 4\,Gyr.  Subsequently the galaxy crosses the GV swiftly in about 230\,Myr.
Galaxy g6.70e12 has a high SFR starting at 1\,Gyr, which then declines gradually. Thus the galaxy enters the GV quite late at about 8\,Gyr, then again at 10\,Gyr. The galaxy continues to form stars while in the GV, leading to a rather long GV crossing time of 1.7\,Gyr.
Galaxy g2.02e13 crosses the GV in 320\,Myr at about 4\,Gyr due to SF quenching. However, a subsequent starburst rejuvenates the galaxy, it then crosses the GV again in about 320\,Myr.
Galaxy g2.37e13 even crosses the GV three times, in 340\,Myr, 330\,Myr and 130\,Myr, due to continued starbursts.
Galaxy g2.79e12 enters the GV quite late at 8\,Gyr and crosses it in 1.3\,Gyr. However, during the last 0.8\,Gyr of its journey through the GV, the galaxy moves parallel and very close to the upper boundary of the GV, meaning a small change in another galactic quantity (e.g. the stellar metallicity or stellar mass) could cause the galaxy to cross in only 0.5\,Gyr.
Galaxy g1.26e12 is subject to continued star formation, which starts to decline after 10\,Gyr. However, the decline is too shallow, thus the galaxy does not cross the GV at all and ends as blue galaxy.

\subsection{Synthetic stellar populations}\label{sec:ssp}

As the star formation history and colour evolution of observed and simulated galaxies is quite complex and influenced by many different physical phenomena (star formation, starbursts, quenching, merging, etc.), we construct synthetic stellar populations (SSP) with a `simple' star formation history and analyze their colour evolution.
We want to see which star formation histories actually lead a galaxy to cross the GV.
These SSPs are constructed by creating star particles with a specific mass, formation time, and metallicity that reproduce a specific star formation history. 

\begin{figure}
	\includegraphics[width=\columnwidth]{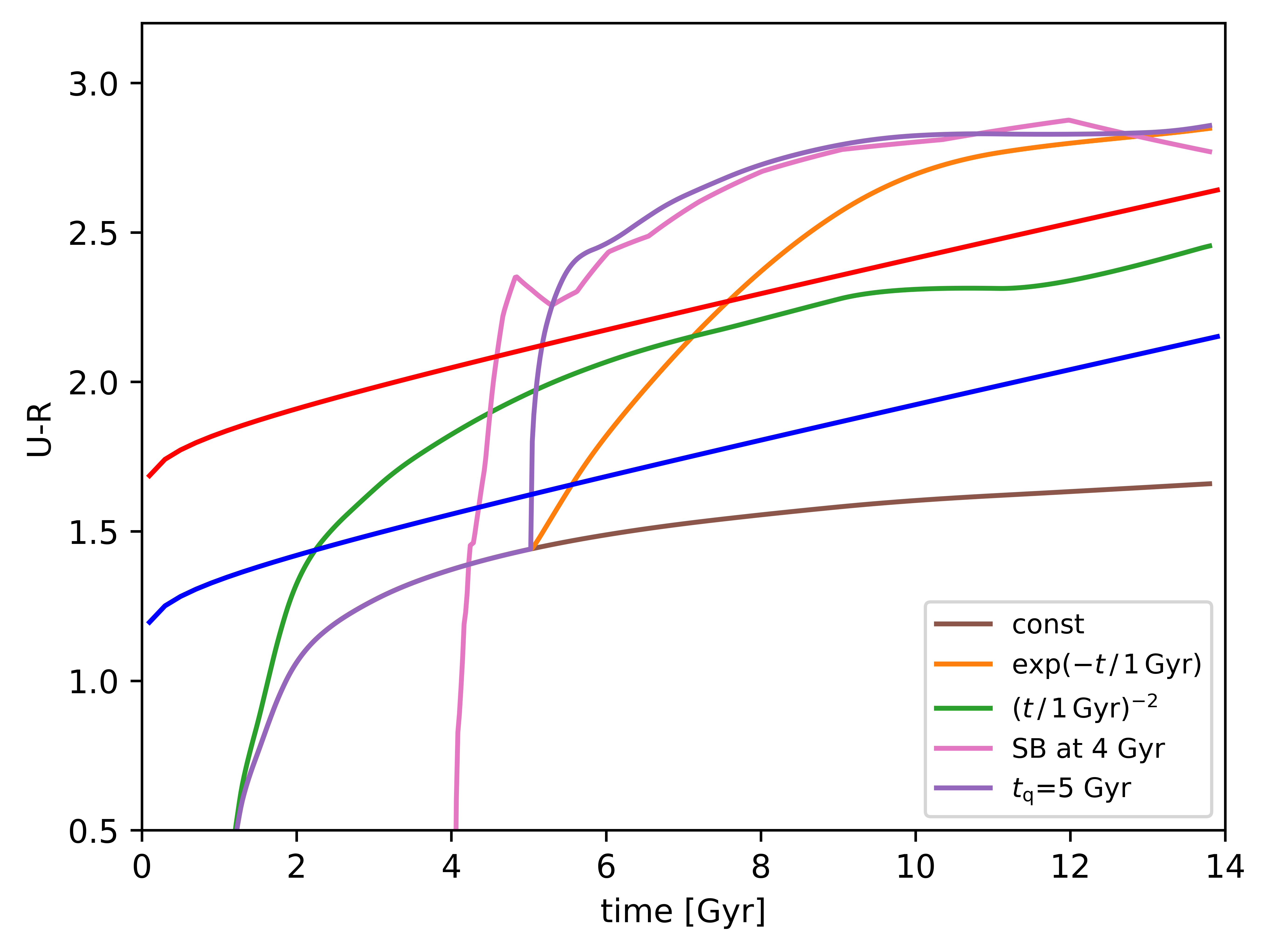}
    \caption{U-R colour versus time for different star formation histories: SFR~$\sim$ const (brown), SFR~$\sim \exp(-t/1\,\mathrm{Gyr})$ (orange), SFR~$\sim t^{-2}$ (green), starburst (i.e. the SFR is a delta function) at $t=4\,\mathrm{Gyr}$ (pink), constant SFR dropping to zero at $t=5\,\mathrm{Gyr}$ (purple).
    The red and blue line are the upper and lower boundary of the GV for the orange SF history.
    The metallicity is $Z=0.02$ for all SF histories.}
    \label{fig:lfile_sfr}
\end{figure}

Fig.~\ref{fig:lfile_sfr} shows the colour evolution for different functions of the SFR (as a function of time $t$) with the metallicity fixed at $Z=0.02$. SSPs with a constant SFR versus time or a power-law SFR $\sim t^{-2}$ never cross the GV within a cosmological time. Only functions with a strong decline, such as the sudden drop of a constant SFR to zero, a starburst, or a SFR $\sim \exp(-t)$, cross the GV, indicating that a quick abrupt change in the SFR, i.e. some sort of SF quenching, is necessary to cross the GV.
We note that the normalization of the SF histories does not change the U-R colour, but it has an effect on the GV boundaries, as those are a function of stellar mass. Thus strictly speaking each of the SF histories presented has its own GV boundaries.
However, the different boundaries are very similar and would be barely distinguishable in the plot; therefore, we only plot the boundaries of the SFR $\sim \exp(-t\,/\,1\,\mathrm{Gyr})$ (orange lines).

\begin{figure*}
	\includegraphics[width=\textwidth]{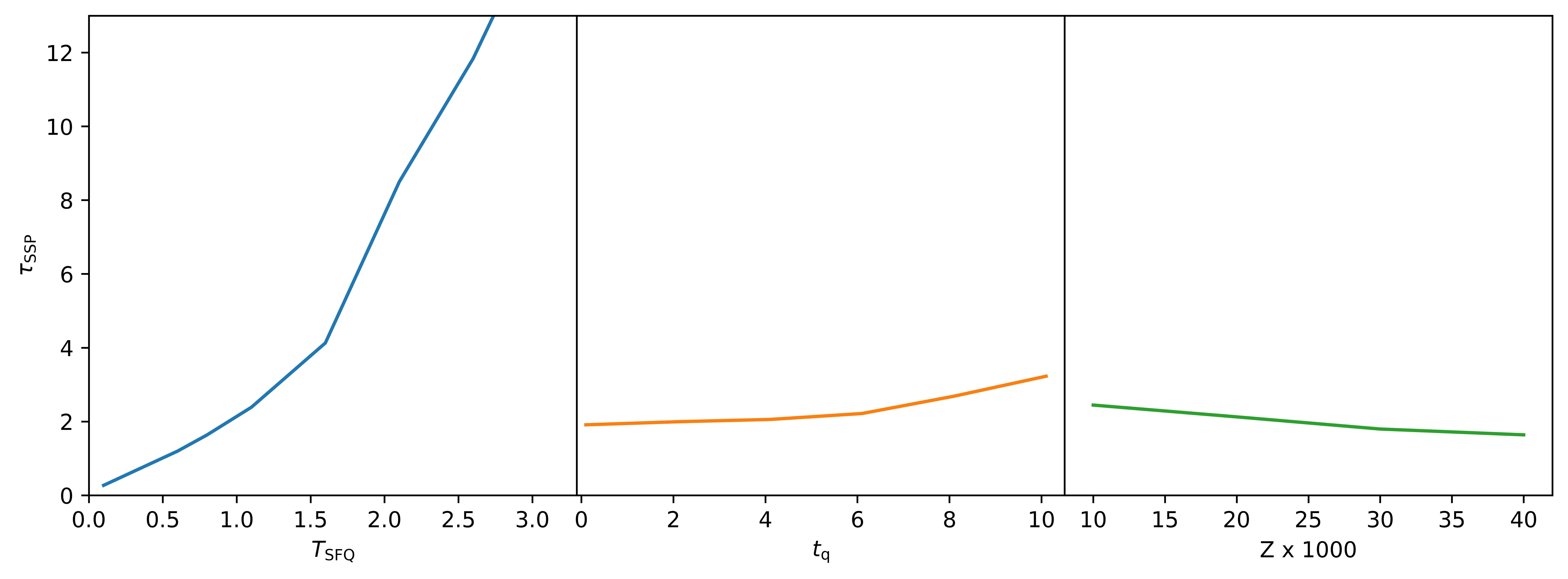}
    \caption{GV crossing time for the synthetic stellar populations, for a constant SFR that changes to $\mathrm{SFR}\sim \exp(-t/T_{\mathrm{SFQ}})$ at $t_{\mathrm{q}}$, as a function of: Left panel: SF quenching timescale $T_{\mathrm{SFQ}}$, middle panel: quenching time $t_{\mathrm{q}}$, right panel: metallicity $Z$.}
    \label{fig:lfile_tgreen}
\end{figure*}

Fig.~\ref{fig:lfile_tgreen} shows the GV crossing time for a SFR that is first constant, and then at a {\it quenching time} $t_{\mathrm{q}}$ changes to an exponentially declining SFR with {\it SF quenching timescale} ($e$-folding timescale) $T_{\mathrm{SFQ}}$ (orange line in Fig.~\ref{fig:lfile_sfr}).
The left panel shows the crossing time for different SF quenching timescales, the middle panel for different quenching times, and the right panel for different metallicities. Quenching time and metallicity have very little effect on the GV crossing time and give values of around 2~Gyr. Only the SF quenching timescale has a large effect on the crossing time, showing variations from 0.2 to more than 12 Gyr. Furthermore, for crossing times up to 2~Gyr, the slope of this function is about two, meaning the crossing time is about twice the SF quenching timescale. 

In previous works \citet{2014_Schawinski_Urry_Simmons} explore exponential declining SFRs with different quenching timescales, which lead the galaxies to take different paths through the GV, with larger quenching times leading to larger crossing times.
Also \citet{2015_Smethurst_Lintott_Simmmons} explore those SFRs, and show that early- and late-type galaxies have different pathways through the GV.

\subsection{Simulated galaxies}\label{sec:sims}

After looking at SSPs, we now investigate the colour evolution and GV crossing times of the NIHAO galaxies and compare them with values from the SSPs introduced in the last subsection.
The GV crossing time of a galaxy consists of three contributions: (i) the age and metallicity of the galaxy's stellar population when it enters the GV (called {\it simplified model}), (ii) star formation while in the GV (called {\it SF-only model}), and (iii) ex-situ stars, i.e. stars that form outside and then enter the galaxy or stars that leave the galaxy (although the latter contribution is very small).
Thus the {\it fiducial crossing time} as seen in the simulations can be expressed as
\begin{equation}
    \tau_{\mathrm{fid}} = \tau_{\mathrm{simp}} + \Delta \tau_{\mathrm{SFON}} + \Delta \tau_{\mathrm{XS}},
    \label{eq:tgreen}
\end{equation}
where $\tau_{\mathrm{simp}}$ is the {\it simplified crossing time}, $\Delta \tau_{\mathrm{SFON}}$ is the {\it SF-only overtime}, and $\Delta \tau_{\mathrm{XS}}$ is the {\it ex-situ overtime}.
We also define the {\it SF-only crossing time} $\tau_{\rmn{SFON}}=\tau_{\rmn{simp}}+\Delta \tau_{\rmn{SFON}}$.
In this subsection, we investigate these three contributions independently from each other: contribution (i) in subsection \ref{sec:tgreen_simple}, (ii) in subsection \ref{sec:sfon}, and (iii) in subsection \ref{sec:exsitu}.
In subsection \ref{sec:combined}, we investigate all contributions combined.

\subsubsection{Simplified model}\label{sec:tgreen_simple}
\begin{figure*}
\includegraphics[width=\textwidth]{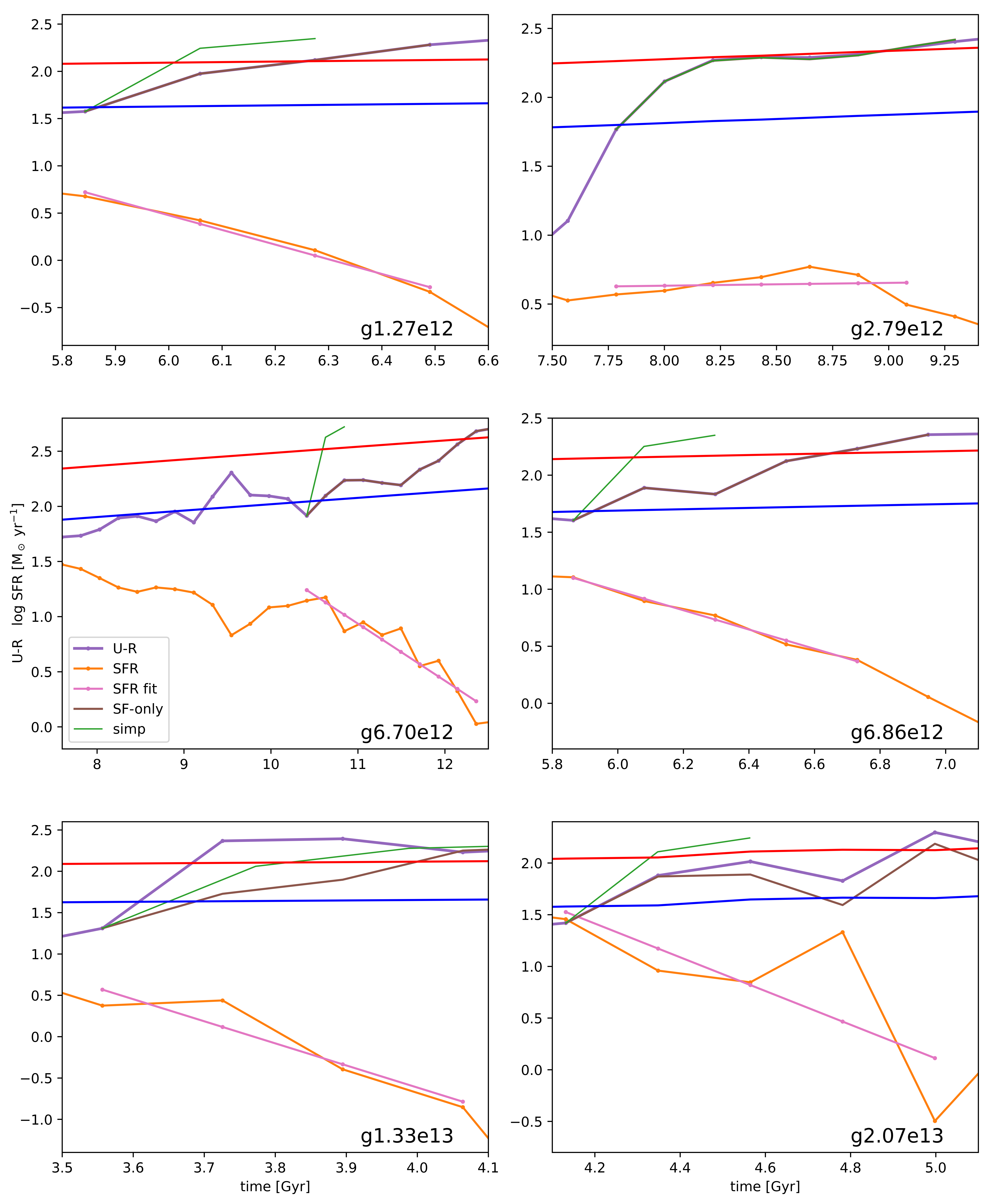}
    \caption{U-R colour (purple) and SFR (orange) versus time for six galaxies. The red and blue line are the upper and lower boundary, respectively, of the GV.
    We also show the colour evolution for the simplified model (green, section \ref{sec:sims}) and the model only considering star formation (but no merger) in the GV (brown, section \ref{sec:sfon}). We furthermore show a linear fit (pink) to the logarithmic SFR in the GV.}
    \label{fig:mag_time_tracks_comp2}
\end{figure*}

For investigating the sole effect of stellar age and metallicity on the GV crossing time, we make some simplifications: after the galaxy has entered the GV, we disregard all stars that subsequently form within the galaxy or those that enter or leave the galaxy. Therefore, the change in colour is only driven by the ageing of the stars already present and not by new stars forming or other stellar populations merging into or leaving the galaxy.
Examples for this evolution are shown in Fig.~\ref{fig:mag_time_tracks_comp2} as green lines: E.g. galaxy g2.79e12 only shows a barely noticeable difference between simple (green) and fiducial (purple) colour evolution leading to only a very small difference of 30~Myr in crossing time. For the galaxy g6.70e12, the difference is much bigger, and the fiducial crossing time of 1.7\,Gyr is reduced to 150\,Myr for the simplified model. Galaxy g1.33e13 is the only one where the simplified crossing time is larger than the fiducial crossing time, caused by a merger that significantly reduces the fiducial crossing time.

\begin{figure*}
	\includegraphics[width=\textwidth]{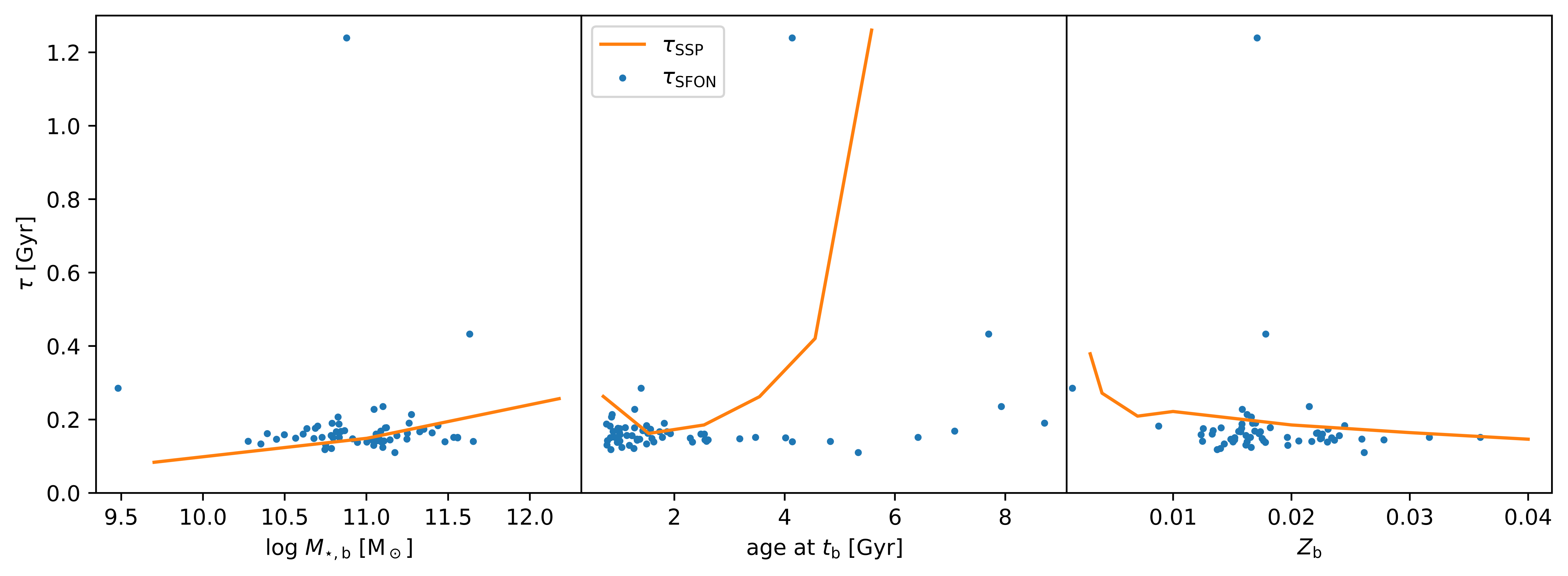}
    \caption{GV crossing time (simplified) as a function of stellar mass (left panel), mean stellar age (middle panel), and the mean stellar metallicity (right panel).
    Blue circles: NIHAO simulations, orange lines: SSP.}
    \label{fig:tg_simp}
\end{figure*}

Fig.~\ref{fig:tg_simp} shows the simplified crossing time as a function of stellar mass (left panel), mean age of the stellar population (middle panel), and the mean stellar metallicity (right panel), with the NIHAO simulations as blue dots and the SSPs as orange lines.
The SSPs have a constant SFR of $60\,\mathrm{M}_{\sun}\,\mathrm{yr}^{-1}$ until $t_{\rm q} = 5\,\mathrm{Gyr}$ and then, are quenched to zero. For the crossing time as function of stellar mass (left panel), we vary the constant SFR between 1 and 300\,$\mathrm{M}_{\sun}\,\mathrm{yr}^{-1}$, and for the crossing time as function of age, we vary the quenching time $t_{\mathrm{q}}$ between 1 and 13\,Gyr.
With stellar mass, the crossing time increases only very slightly for the SSPs; for the NIHAO simulations, this effect is washed out by the scatter of this relation.
With mean stellar age, the crossing time for the SSPs first decreases slightly, then increases with mean stellar age. For the NIHAO simulations, the slight decrease is reproduced; the increase however is only followed by one galaxy with a crossing time of about 1.2\,Gyr. Otherwise for large mean stellar ages, the NIHAO simulations give a constant crossing time of around 200\,Myr. This discrepancy between SSP and NIHAO simulations is due to the model of a constant SFR that is quenched at $t_{\mathrm{q}}$ being a bad approximation for the NIHAO galaxies with large mean stellar ages.
With metallicity, the crossing time for the SSPs is generally decreasing. The NIHAO galaxies mostly follow the slight decrease and give a crossing time of around 200\,Myr.

Generally all three quantities (stellar mass, mean stellar age, and mean stellar metallicity) have a negligible effect on the crossing time. The average crossing time is 180\,Myr.

\subsubsection{SF-only model}\label{sec:sfon}

Star formation within the GV has the effect to extend the time a galaxy spends there. In this subsection, we take a closer look at the SF-only crossing time $\tau_{\rmn{SFON}}$ model that only considers the ageing of the stellar population as it exists at the time the galaxy enters the GV and any subsequent star formation that occurs while traversing the GV, but not stars merging into or out of the galaxy.
Examples for this evolution are shown in Fig.~\ref{fig:mag_time_tracks_comp2} as brown lines: They coincide with the fiducial colour evolution (purple line) for most galaxies; only few show significant deviations due to ex-situ stars, which we will elaborate on in section \ref{sec:exsitu}.

\begin{figure*}
	\includegraphics[width=\textwidth]{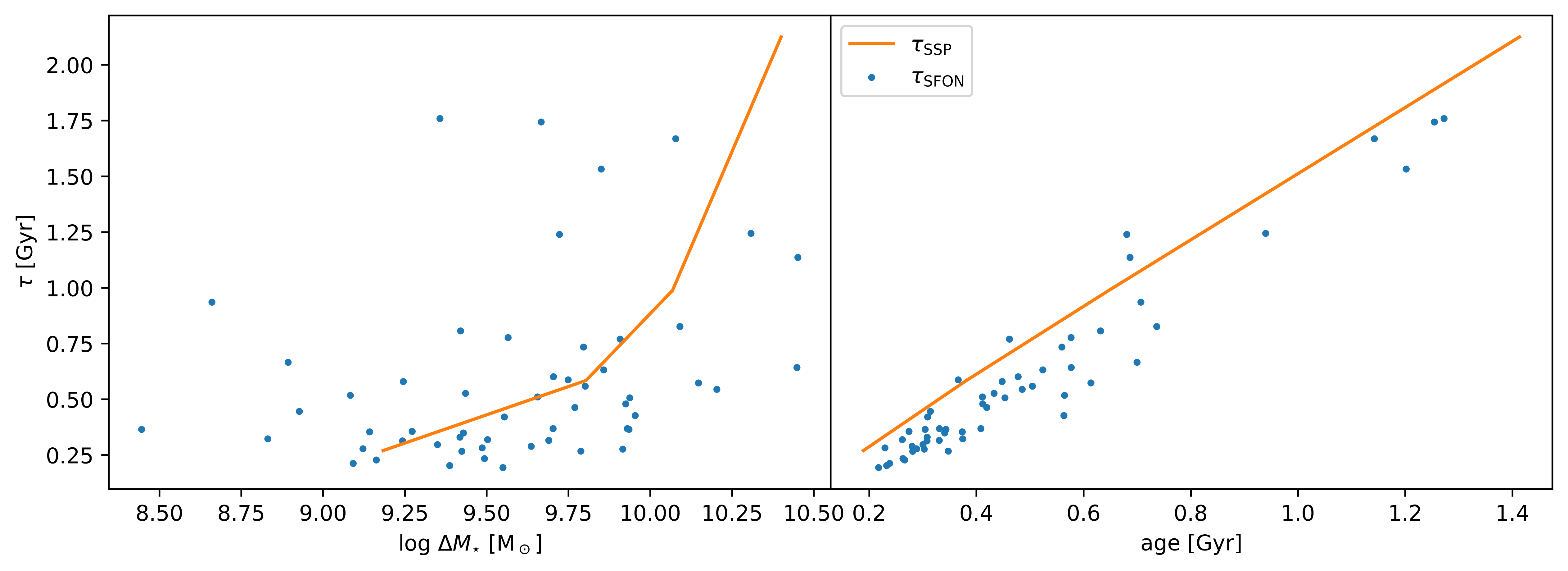}
    \caption{GV crossing time $\tau_{\rmn{SFON}}$ as a function of the mass of the stars that formed in the GV (left panel), and the mean stellar age of the stars that formed in the GV (right panel). Blue circles: NIHAO simulations, orange lines: SSP.}
    \label{fig:tg_sfon}
\end{figure*}

Fig.~\ref{fig:tg_sfon} shows the SF-only crossing time $\tau_{\rmn{SFON}}$ as a function of the mass and the mean stellar age of the stars that form in the GV.
The crossing time increases linearly with both quantities, both for the SSPs and the NIHAO simulations, although the scatter for the function with stellar mass for the NIHAO simulations is quite large. However, a Spearman correlation gives a p-value of $<10^{-4}$, clearly indicating a correlation between these two quantities.

\begin{figure}
	\includegraphics[width=\columnwidth]{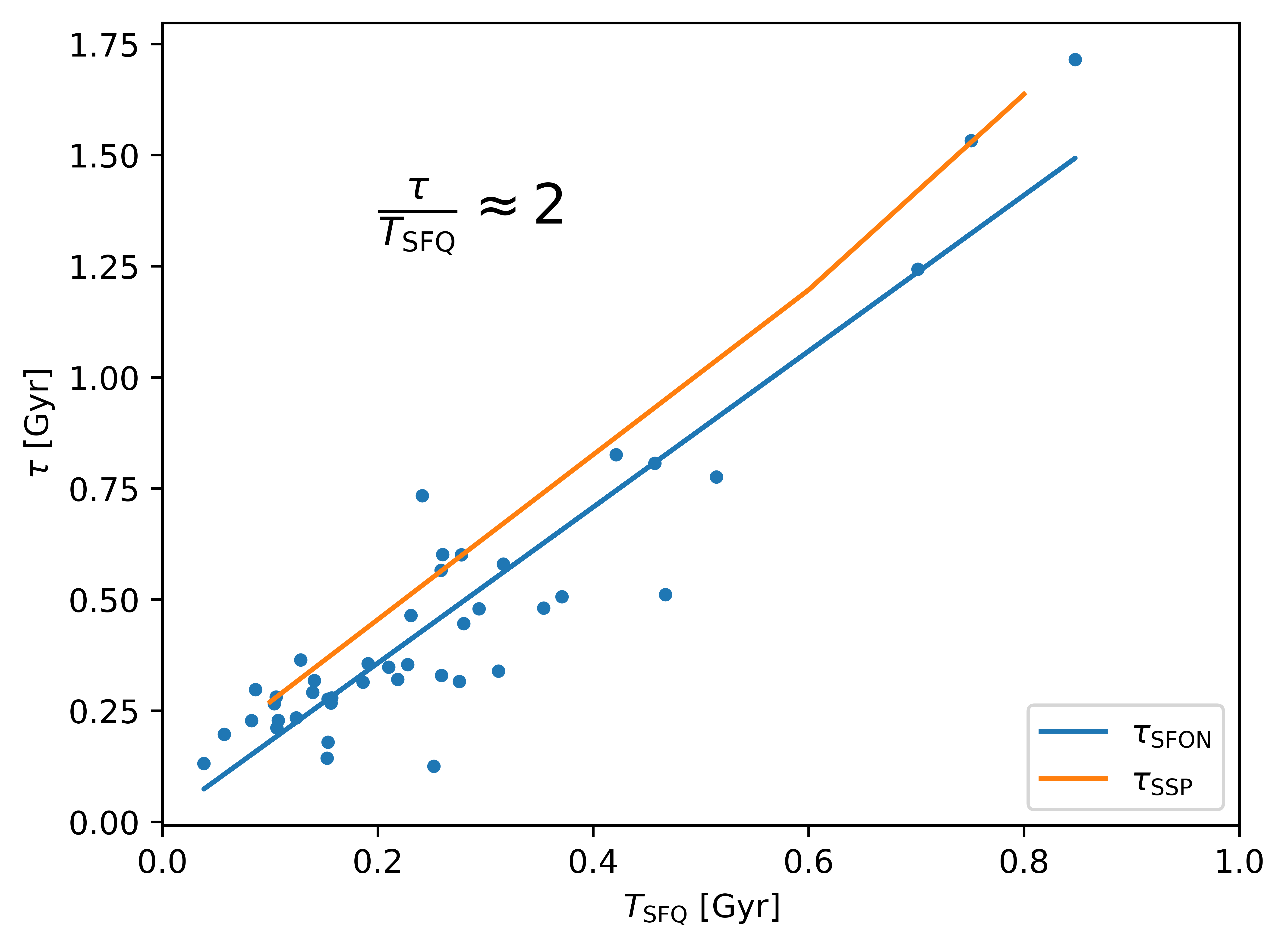}
    \caption{GV crossing time $\tau$ as a function of the SF quenching timescale $T_{\mathrm{SFQ}}$. Blue circles: NIHAO simulations with a monotonically decreasing SFR in the GV, blue line: fit to the NIHAO data, orange line: SSPs (from Fig.\,\ref{fig:lfile_tgreen}).}
    \label{fig:timescales}
\end{figure}

We also want to compare the relation of GV crossing time and SF quenching timescale of the SSPs (Fig.\,\ref{fig:lfile_tgreen}, left panel) with the NIHAO simulations. For this comparison we determine the logarithmic slope of the SFR versus time relation of the NIHAO simulations inside the GV, which is proportional to $-T_{\mathrm{SFQ}}^{-1}$, and thus gives us the SF quenching timescale. This linear fit to the logarithmic SFR is shown in Fig.~\ref{fig:mag_time_tracks_comp2} as pink lines: Many galaxies provide a very good fit (e.g., g1.27e12 and g6.86e12) and a few a bad fit (g2.07e13, due to continued SF in the GV). Galaxy g2.79e12 is the only one with a positive logarithmic slope, thus contradicting the assumption of a declining SFR.
Fig.\,\ref{fig:timescales} then shows the GV crossing time as function of SF quenching timescale, with blue dots for the NIHAO simulations, but only for those that have a monotonically decreasing SFR in the GV. The blue line is a fit to this relation and has a slope of $1.8\pm0.1$. The orange line is the relation for the SSPs from Fig.~\ref{fig:lfile_tgreen}, which has a slope of $1.9\pm0.1$. Thus the GV crossing time is about two times the SF quenching timescale.
\citet{2019_Phillipps_Hopkins_DePropris} already show that the color evolution of a galaxy in only related to the SF quenching timescale.

\subsubsection{Ex-situ stars}\label{sec:exsitu}

\begin{figure*}
	\includegraphics[width=\textwidth]{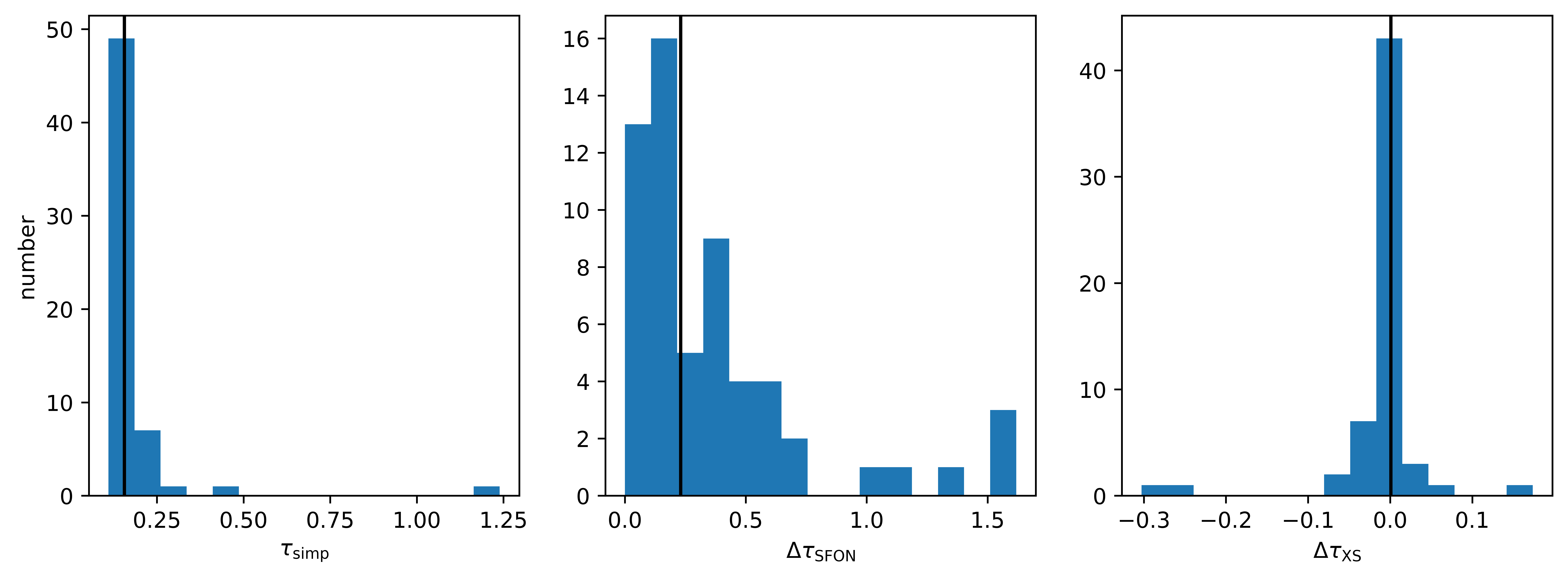}
    \caption{Histograms to the three contributions (eq.\ref{eq:tgreen}) to the GV crossing times for the NIHAO galaxies. The black vertical lines are the averages of the absolute values for the respective contributions.}
    \label{fig:hist_times}
\end{figure*}

In this subsection, we investigate the effect of stars that are part of the galaxy, but have formed elsewhere, i.e. ex-situ stars. These are mostly stars that entered the galaxy through a merger.
Strictly speaking, this effect also considers stars that have left the galaxy, but practically this contribution is very small.
The right panel of Fig.~\ref{fig:hist_times} shows the ex-situ overtime $\Delta \tau_{\rmn{XS}}$, the extra time a galaxy spends in the GV due to ex-situ stars, which we calculate as $\Delta \tau_{\rmn{XS}} = \tau_{\rmn{fid}}-\tau_{\rmn{SFON}}$.
A negative time means the galaxy crosses the GV faster, because the merging stellar population is redder than the host galaxy. A positive time means the galaxy crosses slower, because the merging stellar population is bluer than the host galaxy.
The average of the absolute value of the ex-situ overtime is about 2\,Myr. Thus this effect is generally negligible compared to the contributions from the simplified model and the SF-only model.

Only very few galaxies are significantly affected by ex-situ stars, which is indicated in Fig.~\ref{fig:mag_time_tracks_comp2} by the difference between the fiducial colour evolution (purple lines) and the SF-only model (brown lines).
For galaxy g1.33e13, a merger causes a significant reduction of the crossing time; this is the only galaxy where the simplified crossing time is larger than the fiducial crossing time.
For galaxy g2.07e13, ex-situ stars cause the galaxy to stay in the GV (purple line) instead of leaving the GV and then crossing it (brown line), leading to a much longer fiducial crossing time.

\subsubsection{Fiducial crossing times}\label{sec:combined}

Fig.~\ref{fig:hist_times} shows histograms of the three contributions to the fiducial GV crossing time for the NIHAO galaxies, the black vertical lines denote the average value of the crossing time. These are $\tau_{\mathrm{simp}} = 180\,\mathrm{Myr}$, $\Delta \tau_{\mathrm{SFON}} = 370\,\mathrm{Myr}$, and $\Delta \tau_{\mathrm{XS}} = 2\,\mathrm{Myr}$. Thus the contribution due to ex-situ stars is negligible, 30 per cent of the crossing is due to the ageing of the stars already present when the galaxy enters the GV, and 70 per cent due to SF in the GV.

\subsection{Comparison with other works}\label{sec:comp}

As outlined in section~\ref{sec:intro}, other works report GV crossing times of 1-2\,Gyr and SF quenching times of 0.5-2\,Gyr. Our NIHAO simulations have an average crossing time of 540\,Myr and an average SF quenching time of 280\,Myr. Thus our values are significantly smaller than predicted by other works.
In section~\ref{sec:ssp}, we established that the only quantity with a large effect on the GV crossing time is the SF quenching timescale. Our SSPs with a SF quenching timescale of 1\,Gyr have GV crossing times of 2\,Gyr, which is in agreement with other works.
Thus it seems that the NIHAO galaxies have small GV crossing times, because their SF quenching timescales are small.
In \citet{2019_Blank_Maccio_Dutton}, we established that SF quenching in high-mass galaxies is mostly caused by BH feedback; thus it seems that our models for BH accretion and feedback lead to a very fast quenching of the NIHAO galaxies. Adjustments to these models are needed to reach a more gentle quenching that occurs on a longer timescale.

In section \ref{sec:ssp}, we furthermore established that the ratio of GV crossing time and SF quenching time is always about two.
This ratio does not hold only for our simulations, but also for our SSPs, up to GV crossing times of 3\,Gyr. For even larger crossing times, the relation is steeper, but such large values are not reported by other works.
If the same ratio of GV crossing time and SF quenching time can be found in other works is uncertain, as the reported ranges of GV crossing time and SF quenching times are quite large.

\section{Outlook}\label{sec:outlook}
\begin{figure*}
	\includegraphics[width=0.9\textheight,angle=90]{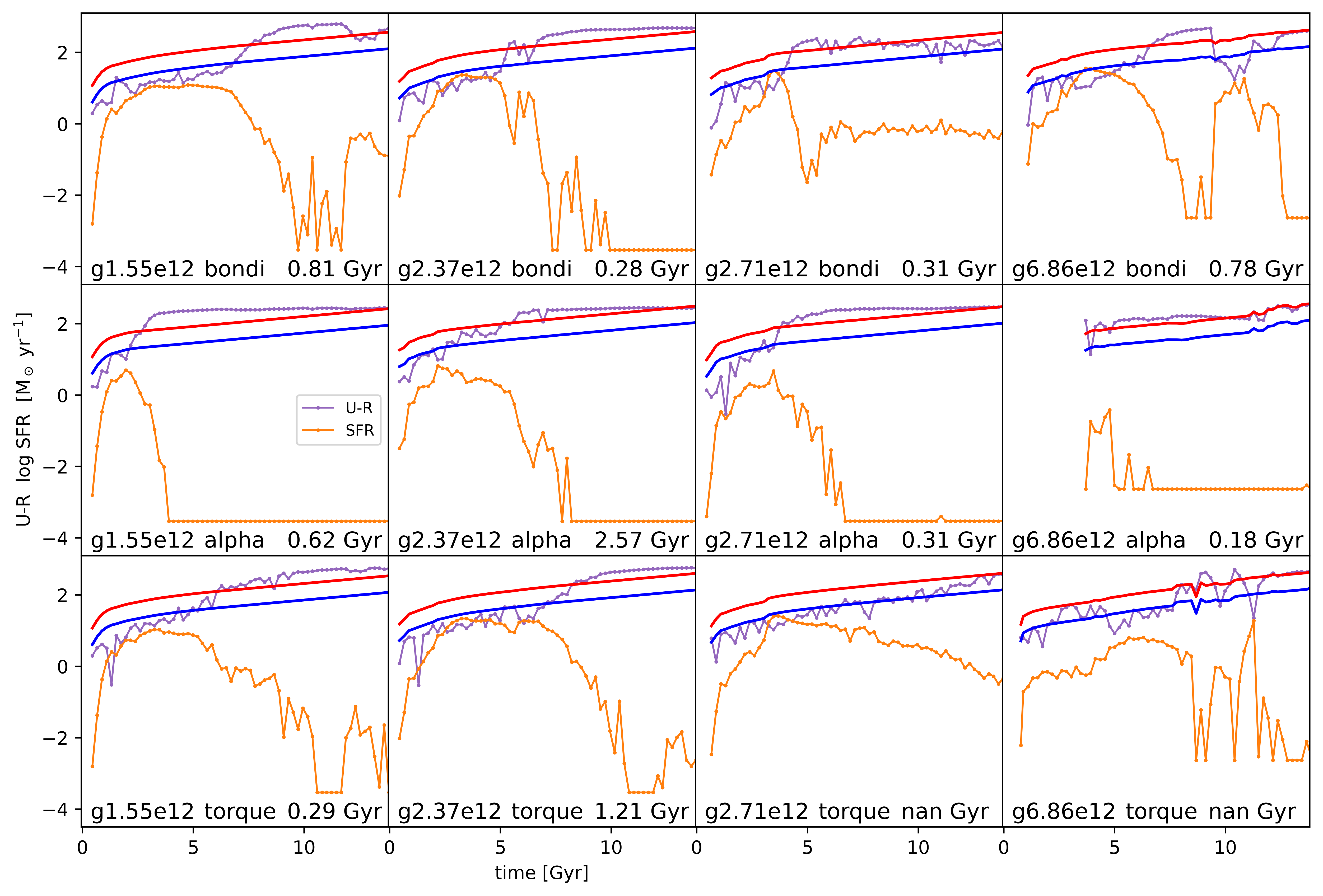}
    \caption{U-R colour (purple) and SFR (orange) versus time for four galaxies (g1.55e12, g2.37e12, g2.71e12, g6.86e12) and three different accretion models (bondi, alpha, torque). The red and blue lines are the upper and lower boundary of the GV, respectively.
    The lower right corner of each subplot shows the GV crossing time, `nan' indicates that the galaxy crosses more or less than one time.}
    \label{fig:mag_time_tracks_nad}
\end{figure*}

In the NIHAO simulations, the BH feedback energy, which quenches SF, is only a function of the BH accretion rate, which is proportional to the BH mass squared (Bondi accretion, see section 2). This model tends to runaway growth as the ever increasing BH mass leads to an ever increasing accretion rate (until the supply of gas is exhausted), causing SF to quench quite rapidly \citep{2019_Blank_Maccio_Dutton}.
Thus changing the model for calculating the BH accretion rate seems the most promising way to increase the GV crossing times of the NIHAO galaxies.
In \citet{2022_Soliman}, we simulate a few NIHAO galaxies with different models for the BH accretion rate. For the {\it torque model} \citep{2011_Hopkins_Quataert}, the BH accretion rate is proportional to the BH mass to the power of 1/6; for the {\it alpha model} \citep{2011_Debuhr_Quataert_Ma}, the BH accretion rate is proportional to the power of -1/2.
In Fig.\,\ref{fig:mag_time_tracks_nad}, we show the colour evolution of four galaxies for three different accretion models (Bondi, alpha, and torque), as well as their GV crossing times.

For the galaxy g1.55e12, the torque model does not yield a higher crossing time, but for the galaxy g2.37e12 the GV crossing time is higher for the torque model than the Bondi model. 
For the galaxy g2.71e12 with torque model star formation gets quenched more gently, indicated by the shallower slope of the SFR. Thus the galaxy spends more time in the GV, but does not cross the GV before the end of the simulation. It might actually cross given a few 100~Myr more time.
The galaxy g6.86e12 crosses the GV three times for the torque model, making comparison with the Bondi model difficult. However, the SFR has a shallower slope compared to the Bondi model.
Thus for the torque model three out of four galaxies show either longer crossing times and/or a shallower decline in the SFR.
These few test simulations look promising in resolving the short SF quenching timescales, and thus the short GV crossing times, but more simulations are necessary to ascertain this trend.
Such a suite of simulations would enable us to investigate the correlations of GV crossing time and other quantities as outlined in section \ref{sec:intro}, such as the gas fraction, ratio of stellar mass and halo mass, morphology, or BH accretion peak time.

\section{Summary}\label{sec:summary}

In our work we study the typical transition time of simulated galaxies through the so-called green valley (GV).
Before moving to complex cosmological simulations, we 
first study in section \ref{sec:ssp} the evolution of synthetic stellar populations (SSPs) that have a variety of simplified star formation histories.
Not surprisingly we find that only synthetic galaxies with a clear truncation (quenching) of star formation activity are able to move from the blue cloud to the red sequence and hence cross the green valley. We find little to no dependence
of this crossing time on the assumed metallicity of
the stellar population or the time at which the quenching event takes place. 
On the other hand, we find a strong correlation between the 
SF quenching timescale (e.g. the $e$-folding timescale for an exponentially declining SF) and the GV crossing time, with a ratio of about a factor of two between the two timescales. 

The results of the simple SSPs model are then confirmed by the study of the fully cosmological hydrodynamical simulations
from the NIHAO suite \citep{2015_Wang_Dutton_Stinson} in section \ref{sec:sims}.
We find three main contributions to the GV crossing time:
(i) the age and metallicity of the galaxy's stellar population when it enters the GV, (ii) star formation while in the GV, and (iii) ex-situ stars, i.e. stars that are accreted via mergers while in the GV.
Contribution (i) is almost always around 200\,Myr, a very low mean stellar metallicity or a very high mean stellar age can in rare cases lead to a much higher GV crossing time. Contribution (ii) is around 370\,Myr, the largest contributor to the GV crossing time. There are some minor positive correlations with stellar mass and age of the stars that form in the GV. Contribution (iii) has a value of about 2\,Myr and is thus negligible for most of the galaxies.

In section \ref{sec:outlook}, we find that different accretion modes for the central BH \citep[from][]{2022_Soliman} can alter the absolute value of the GV crossing time. The classical Bondi-Hoyle accretion \citep{1944_Bondi_Hoyle} leads to too short (compared to observational estimates) crossing times of about 400-500 Myr. Other modes of accretion \citep{2011_Hopkins_Quataert} with a less strong and direct dependence on the BH mass can indeed create
a more gentle SF decline and hence a longer crossing time
of the order of a Gyr, in better agreement with observations.
While different AGN models can give different GV crossing times,  NIHAO galaxies still confirm our findings from section \ref{sec:ssp} that the GV crossing time is about twice the SF quenching timescale, regardless the exact parametrization of BH physics.

Our findings show a direct connection between AGN feedback (and thus BH accretion), SF quenching, and GV crossing time. 
This simple relation can be used to calibrate and test against
observations the many different AGN models implemented in simulations.
More generally, assuming a typical crossing time of about one Gyr inferred from observations \citep[e.g.][]{2018_Bremer_Phillipps_Kelvin,2014_Schawinski_Urry_Simmons,2013_Wetzel_Tinker_Conroy}, our results imply that any mechanism or process aiming to quench star formation, must do so on a typical timescale of about 500 Myr.

\section*{Acknowledgements}

The authors gratefully acknowledge the Gauss Centre for Supercomputing e.V. (www.gauss-centre.eu) for funding this project by providing computing time on the GCS Supercomputer SuperMUC at Leibniz Supercomputing Centre (www.lrz.de).
A part of this research was carried out on the High Performance Computing resources at New York University Abu Dhabi.
We used the software package {\sc pynbody} \citep{pynbody} for our analyses.


\section*{Data Availability}

The data underlying this article will be shared on reasonable request to the corresponding author.



\bibliographystyle{mnras}
\bibliography{library} 






\bsp	
\label{lastpage}
\end{document}